# LARGE-SCALE OPTICAL DIFFRACTION TOMOGRAPHY FOR INSPECTION OF OPTICAL PLASTIC LENSES

## KYOOHYUN KIM, JONGHEE YOON, YONGKEUN PARK[*]

*Department of Physics, Korea Advanced Institute of Science and Technology, 305-701, Daejeon, Republic of Korea*
*Corresponding author: yk.park@kaist.ac.kr*

**Herein is presented an optical diffraction tomography (ODT) technique for measuring 3-D refractive index (RI) maps of optical plastic lenses. A Mach-Zehnder interferometer was used to measure multiple complex optical fields of a plastic lens immersed in RI matching oil, at various rotational orientations. From this, ODT was used to reconstruct a 3-D RI distribution of the plastic lens with unprecedented RI sensitivity ($\Delta n = 4.21 \times 10^{-5}$) and high resolution (12.8 µm). As a demonstration, 3-D RI distributions of a 2-mm-diameter borosilicate sphere and a 5-mm-diameter plastic lens were reconstructed. Defects in the lens, generated by pulsed laser ablation, were also detected using the present method.**

Optical plastic lenses have replaced glass lenses in various fields, particularly in the mobile phone industry, due to their free-form geometry, low weight, high imaging quality, and low production cost. The injection molding process, however, inherently induces molecular orientations of plastic materials during solidification, and may cause inhomogeneous optical properties that can deteriorate the imaging quality of plastic lenses [1, 2].

Inspection of plastic lenses is thus a crucial concern for design, fabrication, and quality control. To this end, various techniques have been developed including surface profilometry [3], modulation transfer function measurements [4], and wavefront measurements using a Shack-Hartmann sensor [5] or interferometry [3]. Previous approaches, unfortunately, did not meet the requirements of simultaneously measuring the 3-D shapes and optical properties of lenses. Surface profilometry only measured the surface shape of lenses. Previous interferometric techniques measure optical-path-length maps, in which RI and thickness information are coupled.

Optical diffraction tomography (ODT) circumvents previous limitations and offers simultaneous and direct measurements of shape and RI maps of a micro lens with improved accuracy [6-8]. ODT has proven to be a useful tool for measuring 3-D refractive index (RI) maps of transparent samples. Using a diffraction tomography reconstruction algorithm, 3-D RI maps of a sample are obtained from multiple 2-D holograms measured at various angles of illumination, or rotations of a sample. ODT has been employed for the inspection of optical elements such as optical fibers and waveguides [9, 10] as well as for the study of pathophysiology of biological cells [11-16].

Here, we present measurements of 3-D RI distribution of plastic lenses using ODT. Using Mach-Zehnder interferometry, multiple 2-D optical fields of a plastic lens were measured at various orientations, from which 3-D RI distribution of the lens was reconstructed. As a proof-of-principle demonstration, we first measured 3-D RI map of solid-glass beads with known RI value, and then applied this to measuring the 3-D RI distribution of plastic lenses with high spatial resolution (12.8 µm) and RI sensitivity ($\Delta n = 4.21 \times 10^{-5}$). Furthermore, we were also able to detect damage to lenses altered by pulsed laser ablation.

To obtain 3-D RI maps of optical plastic lenses, multiple 2-D optical fields of a sample at various rotational orientations were first measured using Mach-Zehnder interferometry [Fig. 1(a)]. A diode-pumped solid-state laser beam ($\lambda$ = 532 nm, 100 mW, Samba, Cobolt, Inc.) was divided into two arms by a beam splitter. In the sample arm path, a sample, attached to a micropipette tip glued by UV-curing optical adhesive (NOA 61, Norland Products Inc.), was placed on a motorized rotating mount (PRM1/MZ8E, Thorlabs Inc.). The sample was submerged in RI-matching oil (Series A, Cargille-Sacher Laboratories Inc.) in a custom-made slide glass chamber [Fig. 1(b)], in order to minimize optical phase delay.

Because lenses are typically at millimeter scale, the optical phase delay of a beam passing through a micro lens significantly increases, even when the RI contrast between a sample and the surrounding medium slightly differs. For example, a slight RI difference of 0.01 will result in optical phase delay greater than 100 rad for a 10-mm-diameter micro lens. For this reason, the RI value of the matching oil was selected cautiously so that optical phase delay of a sample with any orientations was minimized (< $2\pi$ rad). To yield the desired intermediate RI value of the matching oil, two types of RI matching oil were mixed in the necessary volumetric ratio.

The optical field diffracted by a sample was collected by plano-convex lenses in a 4-$f$ telescopic configuration with the additional magnification of 1.33× [L1 – L2 in Fig. 1(a)]. Then, the sample beam interferes with the reference beam at the image plane of the sample, and generates a spatially modulated hologram. For tomographic reconstruction, a motorized mount was used to rotate the sample from 0° to 180° in sequential angular steps of 0.5°. A digital single-lens reflex (DSLR) camera (1100D, Canon Inc., 12 mega pixels with a pitch size of 5.2 µm) was used to record holograms of the sample at various rotational orientations.

From the measured holograms, complex optical fields with various orientations were retrieved using a field retrieval algorithm based on the Fourier transform [17, 18]. As shown in Fig. 1(c), the retrieved optical phase maps of the plastic lens at various orientations have phase values < $2\pi$ rad due to RI matching. According to the Fourier diffraction theorem [6, 7], each 2-D Fourier spectrum of measured optical fields was mapped onto a 3-D Ewald sphere with the corresponding orientation angle of the sample [Fig. 1(d)]. Taking the 3-D inverse Fourier transform of the mapped 3-D Fourier space reconstructs the 3-D RI distribution of the sample. Here, the first Rytov approximation was implemented for ODT, which is valid because the phase gradient of the sample does not vary significantly after RI matching [19].

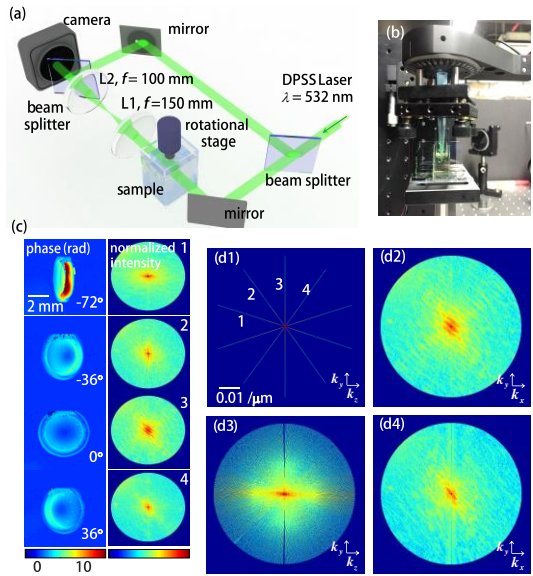

Fig. 1. (a) The experimental setup: L1-2: lenses, (b) Photograph of a sample, attached to a rotation mount, immersed in RI matching oil, (c) Retrieved 2-D phase delay maps (left panel) and corresponding Fourier spectra (right panel) of an optical plastic lens at various rotational orientations, (d1-d2) 3-D Fourier space in the $k_y$-$k_z$ and $k_x$-$k_y$ plane, respectively, reconstructed by the Fourier diffraction theorem with five representative sample orientations indicated in (c), (d3-d4) 3-D Fourier space in the $k_y$-$k_z$ and $k_x$-$k_y$ plane, respectively, reconstructed with full 180° sample orientations.

To prove the validity of the present method for measuring the 3-D RI distribution of large samples, we first measured 3-D RI map of a borosilicate solid-glass sphere of 2-mm diameter (Z273627, Sigma-Aldrich Inc.). The cross-sectional slices of the reconstructed 3-D RI tomogram of the sphere [Fig. 2(a)] clearly show the homogeneous RI distribution inside the sphere, with the correct RI value ($n$ = 1.4734), and the rendered isosurface [Fig. 2(b)] revealed the spherical shape of the sample. For the purpose of visualization, the top region of the sphere was omitted because amorphous UV curing adhesives and a micropipette tip attached to the top showed poor tomographic reconstruction, caused by their high RI contrast against the surrounding medium.

The lateral and axial resolution of the reconstructed tomograms had identical values (12.8 μm) calculated from the maximum spatial frequency in the 3-D Fourier space acquired by the setup. Unlike the illumination scanning approach, which exhibited poorer axial than lateral resolution due to missing information from the limited acceptance angle of a detecting lens, the sample rotation approach used in this work provided uniform axial and lateral resolution [11, 20]. Furthermore, the sample-rotation approach utilized does not require additional normalization processes to fill in the missing information, such as for the iterative non-negativity or total variance minimization processes [21].

The measured RI tomograms exhibited very high RI sensitivity. The standard deviation of the RI difference values in the background region was $\Delta n$ = 4.21 × $10^{-5}$. This unprecedented RI sensitivity originated from the stable optical-field measurement and RI matching between the sample and the medium.

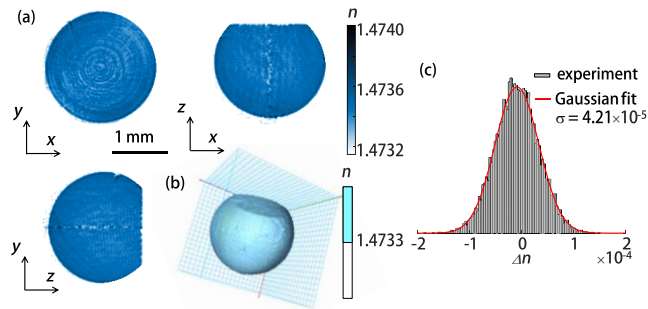

Fig. 2. (a) The cross-sections of the 3-D RI distribution of a borosilicate solid-sphere along the *x-y*, *x-z*, and *y-z* plane, respectively. (b) A 3-D rendered isosurface of the reconstructed RI tomogram of the solid-sphere. (c) The distribution of RI contrast values in the background region. The solid red line indicates a Gaussian fit with the standard deviation of 4.21 × $10^{-5}$.

The present technique was then applied to measure the 3-D RI distribution of plastic lenses. A plastic lens, disassembled from a camera unit in a mobile phone (iPhone 4, Apple Inc.), was submerged in RI matching oil. The *x-z* and *y-z* cross-sectional slices and rendered isosurface of the reconstructed 3-D RI distribution clearly show the aspherical shape of the plastic lens [Fig. 3]. The averaged RI value of the lens was 1.53497 ± 0.00008. As shown in Fig. 3(a), the RI distribution of the lens in the *x-y* plane is inhomogeneous compared to the solid-glass sphere (Fig. 2). This result may imply that the optical inhomogeneity of the plastic lens originated from the molecular orientation of the plastic material during the injection molding process.

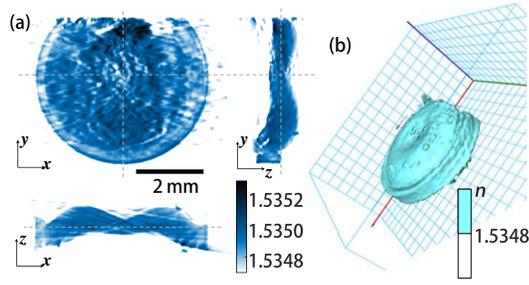

Fig. 3. (a) Cross-sectional slice of the 3-D RI distribution of an optical plastic lens along the *x-y*, *x-z*, and *y-z* plane, (b) A 3-D rendered isosurface of the reconstructed RI tomogram of the plastic lens.

To further demonstrate the applicability of the present method, we measured the 3-D RI distribution of a plastic lens with defects on the surface. The defects on the plastic lens were intentionally generated by ablating the lens surface using a high-power ultrashort pulsed laser beam via nonlinear absorption [22]. To generate defects, we used a two-photon laser scanning microscope (LSM510, Carl Zeiss AG) with a mode-locked Ti-sapphire laser ($\lambda$ = 800 nm, 120 fs, Chameleon, Coherent Inc.). The pulse peak intensity was $0.95 \times 10^{17}$ W/m$^2$, which is strong enough to ablate a plastic lens. The lens was irradiated by femtosecond laser pulses for 168.83 μs over an area of 320 × 320 μm$^2$. Reflection-mode confocal microscopy was used to confirm and locate the defects so generated [Fig. 4(a)].

While the measured 2-D phase map of the damaged lens, in face-on orientation, indicated defects (see green and blue arrows in Fig. 4b), it was difficult to determine the axial location of the defects and whether these defects were surface scratches or internal molecular inhomogeneity. However, the reconstructed 3-D RI distribution and the isosurface image of the lens clearly showed that these defects were located on the surface of the lens [Figs. 4(c)-(d)]. Furthermore, the measured tomogram indicated that the defect regions had RI values identical to that of the surrounding matching oil, suggesting that the difference was caused by the elimination of the surface by the high-powered pulsed laser beam, rather than by modification of the optical properties of the lens. This result clearly demonstrates that this new method could be effective in characterization and optical inspection of optical plastic lenses.

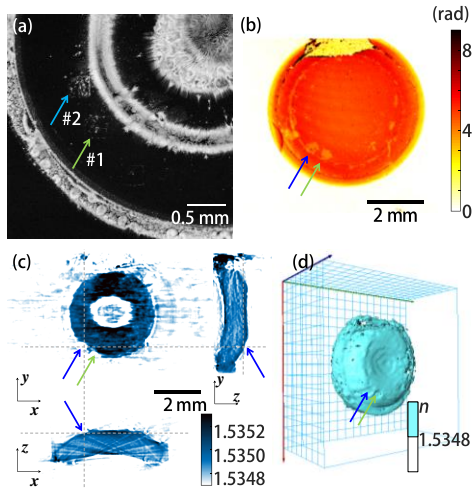

Fig. 4. (a) A reflection-mode confocal fluorescence micrograph of an optical plastic lens with surface defects (ablation by a pulsed laser beam), (b) A 2-D phase map of the plastic lens in face-on orientation, (c) Cross-sections of the 3-D RI distribution of the optical plastic lens along the *x-y*, *x-z*, and *y-z* planes. Dashed lines indicate the relative position of a defect in each plane. (d) Rendered isosurface of a reconstructed RI tomogram of the plastic lens with defects. Green and blue arrows indicate the defects on the surface.

In conclusion, we herein present a novel approach for the 3-D visualization of optical plastic lenses. By measuring multiple 2-D holograms of a lens immersed in RI matching oil at various sample orientations, a 3-D RI distribution was reconstructed with high spatial resolution and RI sensitivity. To demonstrate the precision and capability of the technique, a 2-mm-diameter borosilicate solid-sphere and a 5-mm-diameter optical plastic lens from a commercial smartphone camera were measured. The present paper successfully demonstrates effective measurement of complex 3D shapes (highly aspheric plastic lens) using this technique, and its ability to determine the 3-D positions and types of defects on the lens.

To achieve a field of view large enough to cover the size of lenses (typically 5 mm), we utilized a DSLR camera as an image sensor because it had a large detection area (22.2 × 14.7 mm) with high pixel resolution (12 mega pixels). The resultant spatial resolution of 12.8 μm is high enough to detect optical inhomogeneity [Fig. 3(a)] as well as defects in the lens [Fig. 4(c)]. The total number of optical modes of the present work, calculated as the ratio of image volume size to the diffraction-limited resolution spot volume, was approximately 60 million $\cong$ (50 mm / 12.8 μm)$^3$. This extremely high value of optical modes in ODT has not been reported before, to our knowledge. In the future, the spatial resolution could be further enhanced to reach wavelength resolution using wide-field detectors (e.g., scanning a 1-D line detector or utilizing an image scanner) [23, 24].

The present method provides unprecedented RI sensitivity ($\Delta n = 4.21 \times 10^{-5}$), which enables detection of inhomogeneous spatial distributions of optical materials in the lenses, with high sensitivity. The high RI sensitivity stems from RI matching of thick samples and highly stable phase measurements. Because ODT directly decouples the RI and height information of a sample from measured phase delay maps, thick samples with small phase differences can provide high RI sensitivity. In the present method, however, the use of a coherent laser inevitably causes speckle noise, which deteriorates the 3-D reconstruction quality. Recently, several methods have been introduced to minimize speckle noise in phase measurements [25-29], which could be combined with the present method to further enhance reconstruction quality.

We envision that the present technique could be implemented as an inspection tool for optical plastic lenses, and anticipate that this approach will pave the way for industrial applications of ODT.

**Funding.**

This work was supported by the National Research Foundation (NRF) of Korea (2014K1A3A1A09063027, 2013M3C1A3063046, 2012-M3C1A1-048860, 2014M3C1A3052537), APCTP, and KUSTAR-KAIST project. Kyoohyun Kim is supported by Global Ph.D. Fellowship from NRF.